\documentclass[twocolumn]{article}
\usepackage{sw20lart}

\input tcilatex2.5
\QQQ{Language}{
American English
}

\begin{document}

\author{Hendry I. Elim$^{1\thanks{%
e-mail: elimhi@hotmail.com}}$}
\title{New Integrable Coupled Nonlinear Schr\"{o}dinger Equations}
\date{$^1$\textit{Faculty of Mathematics and Natural Sciences, Department of
Theoretical Physics, Theoretical and Computational Physics Lab., \thinspace
Pattimura University, Ambon, Indonesia}\\
\textbf{ABSTRACT }\\
Two types of integrable coupled nonlinear Schr\"{o}dinger (\textbf{NLS})
equations are derived by using Zakharov-Shabat (\textbf{ZS}) dressing
method.The Lax pairs for the coupled \textbf{NLS} equations are also
\thinspace investigated using the \textbf{ZS} dressing method. These give
new types of the integrable coupled \textbf{NLS} equations with certain
additional terms. Then, the exact solutions of the new types are obtained.
We find that the solution of these new types do not always produce a soliton
solution even they are the kind of the integrable \textbf{NLS} equations.}
\maketitle

\section{\textbf{Introduction}}

One of the most remarkable discoveries in soliton theory is the inverse
scattering method (\textbf{ISM}). It is well-known that there are several
types of the coupled nonlinear Schr\"{o}dinger (\textbf{NLS}) equations
whose complete integrability can prove by some methods in the \textbf{ISM }%
such as\textbf{\ AKNS }method \cite{Kaup}, Wadati method, \textbf{ZS}
dressing method, and so on $\left[ 2-5\right] $. In this paper, we introduce
two new types of the integrable coupled \textbf{NLS} equations derived using 
\textbf{ZS} dressing method. The first type of the equations is 
\begin{eqnarray*}
iu_t\,+\,\chi u_{_{xx}}\,\mp 2\mu \left( \frac{\left| u\right| ^2+\left|
w\right| ^2}{\left| u\right| ^2\left| w\right| ^2}\right) u &=&R_1\, \\
iw_t\,+\,\chi w_{_{xx}}\,\mp 2\mu \left( \frac{\left| u\right| ^2+\left|
w\right| ^2}{\left| u\right| ^2\left| w\right| ^2}\right) w &=&R_2\,,
\end{eqnarray*}
\begin{equation}
\tag{1.1}
\end{equation}
where the perturbative terms $R_1$ and $R_2$, and the real parameters $\chi $%
, and $\mu $ are,respectively, defined as follows 
\begin{equation}
R_1=\frac{2\chi u_{_x}^2}u,  \tag{1.2a}
\end{equation}
\begin{equation}
R_2=\frac{2\chi w_{_x}^2}w,  \tag{1.2b}
\end{equation}
\begin{equation}
\chi =\frac{\alpha +\delta }{\alpha \left( \alpha -\delta \right) }, 
\tag{1.2c}
\end{equation}
and 
\begin{equation}
\mu =\frac{\alpha ^2-\delta ^2}{\alpha ^2\delta }.  \tag{1.2d}
\end{equation}
The second type, 
\begin{eqnarray*}
iu_t\,+\,\chi u_{_{xx}}\,\mp \left( \frac{2\mu \left( \left| u\right|
^2+\left| w\right| ^2\right) u}{1+s\left( \left| u\right| ^2+\left| w\right|
^2\right) }\right)  &=&Q_1\, \\
iw_t\,+\,\chi w_{_{xx}}\,\mp \left( \frac{2\mu \left( \left| u\right|
^2+\left| w\right| ^2\right) w}{1+s\left( \left| u\right| ^2+\left| w\right|
^2\right) }\right)  &=&Q_2\,,
\end{eqnarray*}
\begin{equation}
\tag{1.3}
\end{equation}
is a coupled \textbf{NLS} equation with complicated additional terms ($Q_1$
and $Q_2$) defined as follows 
\begin{equation}
Q_1=Q_{11}+Q_{12}+Q_{13}+Q_{14},  \tag{1.4a}
\end{equation}
and 
\begin{equation}
Q_2=Q_{21}+Q_{22}+Q_{23}+Q_{24},  \tag{1.4b}
\end{equation}
where 
\begin{equation}
Q_{11}=i\left( \frac s2\right) \left( \frac{\left( \left| u\right| ^2+\left|
w\right| ^2\right) _t}{1+s\left( \left| u\right| ^2+\left| w\right|
^2\right) }\right) u,  \tag{1.5a}
\end{equation}
\begin{equation}
Q_{12}=\left( s\chi \right) \left( \frac{\left( \left| u\right| ^2+\left|
w\right| ^2\right) _x}{1+s\left( \left| u\right| ^2+\left| w\right|
^2\right) }\right) u_x,  \tag{1.5b}
\end{equation}
\begin{equation}
Q_{13}=Q_{23}=\left( \frac{s\chi }2\right) \left( \frac{\left( \left|
u\right| ^2+\left| w\right| ^2\right) _{xx}}{1+s\left( \left| u\right|
^2+\left| w\right| ^2\right) }\right) ,  \tag{1.5c}
\end{equation}
\begin{equation}
Q_{14}=-\left( \frac{3s^2\chi }4\right) \left( \frac{\left( \left( \left|
u\right| ^2+\left| w\right| ^2\right) _x\right) ^2}{\left( 1+s\left( \left|
u\right| ^2+\left| w\right| ^2\right) \right) ^2}\right) u,  \tag{1.5d}
\end{equation}
\begin{equation}
Q_{21}=i\left( \frac s2\right) \left( \frac{\left( \left| u\right| ^2+\left|
w\right| ^2\right) _t}{1+s\left( \left| u\right| ^2+\left| w\right|
^2\right) }\right) w,  \tag{1.5e}
\end{equation}
\begin{equation}
Q_{22}=\left( s\chi \right) \left( \frac{\left( \left| u\right| ^2+\left|
w\right| ^2\right) _x}{1+s\left( \left| u\right| ^2+\left| w\right|
^2\right) }\right) w_x,  \tag{1.5f}
\end{equation}
and 
\begin{equation}
Q_{24}=-\left( \frac{3s^2\chi }4\right) \left( \frac{\left( \left( \left|
u\right| ^2+\left| w\right| ^2\right) _x\right) ^2}{\left( 1+s\left( \left|
u\right| ^2+\left| w\right| ^2\right) \right) ^2}\right) w.  \tag{1.5g}
\end{equation}
The parameter $s$ is an effective saturation parameter. If we put $%
s\rightarrow 0$, then the system ($1.3$) reduces to the integrable Manakov
equation $\left[ 6\right] $: 
\begin{eqnarray}
iu_t\,+\,\chi u_{_{xx}}\,\mp 2\mu \left( \left| u\right| ^2+\left| w\right|
^2\right) u &=&0  \nonumber \\
iw_t\,+\,\chi w_{_{xx}}\,\mp 2\mu \left( \left| u\right| ^2+\left| w\right|
^2\right) w &=&0,\,  \tag{1.6}
\end{eqnarray}
Equation ($1.3$) can be identified as the similar equation with that which
has been derived by Christodoulides \textit{et al.} $\left[ 7\right] $, and
it is described by a system of two coupled nonlinear equations for the
normalized beam envelopes, $u\left( x,t\right) $ and $w\left( x,t\right) $ $%
\left[ 8\right] $. However, in eq.($1.3$), we have extended the additional
terms appeared in the related equation provided by Christodoulides \textit{%
et al.} into the complicated terms ($Q_1$ and $Q_2$).

The present paper consists of the following. In section 2, we perform the 
\textbf{ZS} dressing method for the general \textbf{NLS} equations. In
section 3, we propose and derive two types of the integrable coupled \textbf{%
NLS} equations using the \textbf{ZS} method. In section \textbf{4}, we solve
the solutions of the coupled \textbf{NLS} equations. The last section,
section 5\textbf{,} is devoted to the concluding remarks.

\section{\textbf{ZS Dressing Method for the General Coupled NLS Equations}}

\subsection{\textbf{The Integral Operators of The ZS Dressing Method}}

In 1974, Zakharov and Shabat generalised the Lax method using their \textbf{%
ZS} dressing method. In this section, we will review how the \textbf{ZS}
dressing method works on the general \textbf{NLS} equations and will also
show the connection with the work of Lax soon, but first we introduce three
integral operators $\left[ 9\right] $. Let, in general, $F\left( x,z\right) $
and $k_{\pm }\left( x,z\right) $ be $N$x$N$ matrices where 
\begin{equation}
k_{+}\left( x,z\right) =0,\,\,\,\,\text{if}\,\,\,z<x,  \tag{2.1a}
\end{equation}
\begin{equation}
k_{-}\left( x,z\right) =0,\,\,\,\,\text{if}\,\,\,z>x,  \tag{2.1b}
\end{equation}
and let $\psi \left( x\right) $ be an $N$-vector. The integral operators $%
\Phi _F$ and $\Phi _{\pm }$ on $\psi $ are defined by 
\begin{equation}
\Phi _F\left( \psi \right) \,=\,\int\limits_{-\infty }^\infty F\left(
x,z\right) \psi \left( z\right) dz  \tag{2.2}
\end{equation}
for all integrable $\psi $, similarly 
\begin{equation}
\Phi _{\pm }\left( \psi \right) \,=\,\int\limits_{-\infty }^\infty k_{\pm
}\left( x,z\right) \psi \left( z\right) dz,  \tag{2.3}
\end{equation}
so that 
\begin{equation}
\Phi _{+}\,=\,\int\limits_x^\infty k_{+}\left( x,z\right) dz,  \tag{2.4a}
\end{equation}
and 
\begin{equation}
\Phi _{-}\,=\,\int\limits_{-\infty }^xk_{-}\left( x,z\right) dz.  \tag{2.4b}
\end{equation}
We suppose that $\Phi _F$ and $\Phi _{\pm }$ are related by the operator
identity 
\begin{equation}
\left( I\,+\,\Phi _{+}\right) \,\left( I\,+\,\Phi _F\right) \,=\left(
I\,+\,\Phi _{-}\right) ,  \tag{2.5}
\end{equation}
where we assume that $\left( I\,+\,\Phi _{+}\right) $ is\textit{\ invertible}
so that 
\begin{equation}
\left( I\,+\,\Phi _F\right) \,=\left( I\,+\,\Phi _{+}\right) ^{-1}\left(
I\,+\,\Phi _{-}\right) ,  \tag{2.6}
\end{equation}
i.e. the operator $\left( I\,+\,\Phi _F\right) $ is factorisable and $I$ ,as
usual, is the unit matrix.

The identity $\left( 2.5\right) $, on $\psi $, can be written as 
\begin{eqnarray}
&&\int\limits_x^\infty k_{+}\left( x,z\right) \psi \left( z\right)
dz+\int\limits_{-\infty }^\infty F\left( x,z\right) \psi \left( z\right) dz\,
\nonumber \\
&&+\int\limits_x^\infty k_{+}\left( x,z\right) \left( \int\limits_{-\infty
}^\infty F\left( z,y\right) \psi \left( y\right) dy\right) dz\,  \nonumber \\
&=&\int\limits_x^\infty k_{-}\left( x,z\right) \psi \left( z\right) dz, 
\tag{2.7}
\end{eqnarray}
and if we require that $\psi \left( z\right) =0$ for $z<x$ then the
right-hand side is zero. Furthermore, the double integral may be expressed
as 
\[
\int\limits_{-\infty }^\infty \int\limits_x^\infty k_{+}\left( x,y\right)
F\left( y,z\right) \psi \left( z\right) dydz,
\]
where the 'dummy' variables have been relabelled by interchanging $y$ and $z$%
. Thus equation $\left( 2.7\right) $ becomes 
\[
0=\int\limits_x^\infty \left[ k_{+}\left( x,z\right) +F\left( x,z\right)
\right] \psi \left( z\right)
dz\,\,\,\,\,\,\,\,\,\,\,\,\,\,\,\,\,\,\,\,\,\,\,\,\,\,\,\,\,\,\,
\]
\begin{equation}
+\int\limits_x^\infty \int\limits_{-\infty }^\infty \left( k_{+}\left(
x,y\right) F\left( y,z\right) \right) \psi \left( z\right) dy,  \tag{2.8}
\end{equation}
(note the use here of eq.$(2.1a)$ for all $\psi \left( z\right) $).Therefore
\[
\]
\[
0=k_{+}\left( x,z\right) +\,F\left( x,z\right)
\,\,\,\,\,\,\,\,\,\,\,\,\,\,\,\,\,\,\,\,\,\,\,\,\,\,\,\,\,\,\,\,\,\,\,\,\,\,%
\,\,\,\,\,\,\,\,\,\,\,\,\,\,\,\,\,\,\,\,\,\,\,\,\,\,\,\,\,\,\,\,\,\,\,\,\,\,%
\,\,\,\,\,\,\,\,\,\,\,\,\,\,\,\,\,\,\,\,\,\,\,\,\,
\]
\begin{equation}
+\int\limits_t^\infty k_{+}\left( x,y\right) F\left( y,z\right) dy,\,\,\,\,%
\text{\thinspace \thinspace for}\,\,z\,>\,x\,\,.  \tag{2.9}
\end{equation}
Equation $\left( 2.9\right) $ is the matrix Marchenko equation for $%
k_{+}\left( x,z\right) $. Similarly, if we consider $z\,<\,x$ in eq.$(2.9)$,
it can be shown that 
\begin{equation}
k_{-}\left( x,z\right) =\,F\left( x,z\right) +\int\limits_t^\infty
k_{+}\left( x,y\right) F\left( y,z\right) dy\,,  \tag{2.10}
\end{equation}
which defines $k_{-}\left( x,z\right) $ in terms of $k_{+}$ and $F$. At this
stage we have not restricted the choice of $F$, and this we will next do for
some \textbf{NLS} equations.

\subsection{\textbf{The differential Operators of the ZS Dressing Method}}

In common with our previous analyses, we extend the definitions of the
matrices $k_{\pm }$ and $F$, and the vector $\psi $, so that they all may
now depend upon auxiliary variable e.g. $t$, $y$. We shall describe the
evolution of $k_{\pm }$ and $F$ (in $t,y$), and hence relate them to certain
evolution equations by introducing appropriate (linear) differential
oprators. We define the $N$x$N$ matrix differential operator $\Delta
_0^{\left( i\right) }$, $i=1,2$ on $\psi \left( x;t,y\right) $ which has
only constant coefficients and which commutes with the integral operator $%
\Phi _F$, i.e. 
\begin{equation}
\left[ \Delta _0^{\left( i\right) },\Phi _F\right] =\Delta _0^{\left(
i\right) }\Phi _F-\Phi _F\Delta _0^{\left( i\right) }=0.  \tag{2.11}
\end{equation}
Note that in term $\Phi _F\Delta _0^{\left( i\right) }$ , $\Delta _0^{\left(
i\right) }$ operates on $\psi \left( x;t,y\right) $ first and then this is
evaluated on $x=z$ for the application of the operator $\Phi _F$. Further,
we introduce an associated differential operator, $\Delta ^{\left( i\right)
} $, which is defined by the operator identity 
\begin{equation}
\Delta ^{\left( i\right) }\left( I\,+\,\Phi _{+}\right) \,=\,\left(
I\,+\,\Phi _{+}\right) \Delta _0^{\left( i\right) },\,\,\,i=1,2.  \tag{2.12}
\end{equation}
It can be shown that eq.$\left( 2.12\right) $ also holds if $\Phi _{+}$ is
replaced by $\Phi _{-}$:

\textit{Theorem }1. If operator $\left( I\,+\,\Phi _F\right) $ commutes with
the differential operator $\Delta _0^{\left( i\right) }$, and is invertible
then both its Volterra factors $\left( I\,+\,\Phi _{\pm }\right) $ transform 
$\Delta _0^{\left( i\right) }$ into one and the same operator $\Delta
^{\left( i\right) }$.

\textit{Proof }$\left( of\,\,theorem1\right) $. 
\begin{eqnarray}
\Delta ^{\left( i\right) }\, &=&\left( I\,+\,\Phi _{+}\right) \,\Delta
_0^{\left( i\right) }\left( I\,+\,\Phi _{+}\right) ^{-1}  \nonumber \\
&=&\left\{ 
\begin{array}{c}
\left( \left( I\,+\,\Phi _{-}\right) \left( I\,+\,\Phi _F\right)
^{-1}\right) \,\,\,\,\,\,\,\,\,\,\,\,\,\,\,\,\,\,\, \\ 
\,\,\,\,\text{x}\left( \Delta _0^{\left( i\right) }\left( I\,+\,\Phi
_F\right) \left( I\,+\,\Phi _{-}\right) ^{-1}\right)
\end{array}
\right\}  \nonumber \\
&=&\left( I\,+\,\Phi _{-}\right) \,\Delta _0^{\left( i\right) }\left(
I\,+\,\Phi _{-}\right) ^{-1}.  \tag{2.13}
\end{eqnarray}
The operator $\Delta _0^{\left( i\right) }$ is sometimes referred to as '%
\textbf{undressed}', and $\Delta ^{\left( i\right) }$ as the '\textbf{dressed%
}' operator.

Before the general development, we shall consider an example which will
illuminate the meaning of equations $\left( 2.11\right) $ and $\left(
2.12\right) $. Let 
\begin{equation}
\Delta _0^{(1)}\,=\,I\left( i\alpha \frac \partial {\partial t}\,-\,\frac{%
\partial ^2}{\partial x^2}\right) ,  \tag{2.14}
\end{equation}
where $\alpha \,\,$is an arbitrary real value, and $I$ is the $3$x$3$ unit
matrix. Thus eq.$\left( 2.11\right) $, when operated on $\psi \left(
x;t\right) $, becomes 
\begin{eqnarray*}
0 &=&\left( i\alpha \frac \partial {\partial t}\,-\,\frac{\partial ^2}{%
\partial x^2}\right) \int\limits_{-\infty }^\infty F\left( x,z;t\right) \psi
\left( z;t\right) dz \\
&&-\int\limits_{-\infty }^\infty F\left( x,z;t\right) \left( i\alpha \frac
\partial {\partial t}\,-\,\frac{\partial ^2}{\partial z^2}\right) \psi
\left( z;t\right) dz.
\end{eqnarray*}
\begin{equation}
\tag{2.15}
\end{equation}
After integration by parts, it follows that 
\begin{equation}
\int\limits_{-\infty }^\infty F\psi _{zz}dz=\int\limits_{-\infty }^\infty
F_{zz}\psi dz,  \tag{2.16}
\end{equation}
for all bounded continuously twice differentiable $\psi $ provided $\psi $, $%
\psi _z\rightarrow 0$ as $\left| z\right| \rightarrow \infty $. Hence eq.$%
\left( 2.15\right) $ can be written as 
\begin{equation}
\int\limits_{-\infty }^\infty \left( i\alpha F_t-F_{xx}+F_{zz}\right) \psi
\,dz=0,  \tag{2.17}
\end{equation}
and therefore eq.$\left( 2.11\right) $ (for $i=1$) is an operator identity
only if $F\left( x,z;t\right) $ satisfies 
\begin{equation}
i\alpha F_t-F_{xx}+F_{zz}=0.  \tag{2.18}
\end{equation}

The associated operator $\Delta ^{\left( i\right) }$ is now obtained from eq.%
$\left( 2.12\right) \,$(for $i=1$), 
\begin{eqnarray*}
&&\Delta ^{\left( 1\right) }\left\{ \psi \left( x;t\right)
+\int\limits_x^\infty k_{+}\left( x,z;t\right) \psi \left( z;t\right)
dz\right\}  \\
&=&I\left( i\alpha \psi _t-\psi _{xx}\right)  \\
&&+\int\limits_x^\infty k_{+}\left( x,z;t\right) \left( i\alpha \frac
\partial {\partial t}\,-\,\frac{\partial ^2}{\partial z^2}\right) \psi
\left( z;t\right) dz.
\end{eqnarray*}
\begin{equation}
\tag{2.19}
\end{equation}
Again, we integrate by parts to find 
\begin{equation}
\int\limits_x^\infty k_{+}\psi _{zz}dz=-\widehat{k}_{+}\psi _x+\widehat{k}%
_{+_z}\psi +\int\limits_x^\infty k_{+_{zz}}\psi dz,  \tag{2.20}
\end{equation}
where $\widehat{k}_{+}=k_{+}\left( x,x;t\right) $, and we assume that $k_{+}$%
, $k_{+_z}\rightarrow 0$ as $z\rightarrow +\infty $. It is now convenient to
set 
\begin{equation}
\Delta ^{\left( i\right) }=\Delta _0^{(i)}+V_i,\,\,i=1,2\,\,\,\,\,, 
\tag{2.21}
\end{equation}
so that eq.$\left( 2.19\right) $ becomes (for instance, $i=1$) 
\begin{eqnarray}
&&V_1\left( \psi +\int\limits_x^\infty k_{+}\psi dz\right) +i\alpha
\int\limits_x^\infty \left( k_{+_t}\psi +k_{+}\psi _t\right) dz  \nonumber \\
&&+\widehat{k}_{+}\psi _x+\widehat{k}_{+_x}\psi -\int\limits_x^\infty \left(
k_{+_{xx}}-k_{+_{zz}}\right) \psi dz  \nonumber \\
&=&i\alpha \int\limits_x^\infty k_{+}\psi _tdz+\widehat{k}_{+}\psi _x-%
\widehat{k}_{+_z}\psi -\psi \frac d{dx}\widehat{k}_{+},  \tag{2.22}
\end{eqnarray}
or 
\begin{eqnarray}
0 &=&\left( V_1+2\frac d{dx}\widehat{k}_{+}\right) \psi
+V_1\int\limits_x^\infty k_{+}\psi dz  \nonumber \\
&&+\int\limits_x^\infty \left( i\alpha k_{+_t}-k_{+_{xx}}+k_{+_{zz}}\right)
\psi dz,  \tag{2.23}
\end{eqnarray}
since $\frac d{dx}\widehat{k}_{+}=\widehat{k}_{+_x}+\widehat{k}_{+_z}$, the
total derivative in $x$. Hence if this equation is valid for all continuous $%
\psi $, then 
\begin{equation}
V_1\left( x;t\right) =-2\frac d{dx}\widehat{k}_{+}=-2\left( \widehat{k}%
_{+_x}+\widehat{k}_{+_z}\right) ,  \tag{2.24}
\end{equation}
(so that $V_1$ is of degree zero), and $k_{+}\left( x,z;t\right) $ satisfies 
\begin{equation}
i\alpha k_{+_t}-k_{+_{xx}}+k_{+_{zz}}+V_1k_{+}=0.  \tag{2.25}
\end{equation}

On the other hand, we choose the undressed operator $\Delta _0^{(2)}$ as
follows 
\begin{equation}
\Delta _0^{(2)}\,=\,\left( 
\begin{tabular}{lll}
$\alpha $ & $0$ & $0$ \\ 
$0$ & $\delta $ & $0$ \\ 
$0$ & $0$ & $\delta $%
\end{tabular}
\right) \frac \partial {\partial x},  \tag{2.26}
\end{equation}
(where $\alpha \,\,$and $\delta \,$ are arbitrary real values), and then
substitute it into eq.$\left( 2.12\right) $ : $\Delta ^{\left( 2\right)
}\left( I\,+\,\Phi _{+}\right) \,=\,\left( I\,+\,\Phi _{+}\right) \Delta
_0^{\left( 2\right) }$, we obtain 
\begin{eqnarray}
&&\left( 
\begin{tabular}{lll}
$\alpha $ & $0$ & $0$ \\ 
$0$ & $\delta $ & $0$ \\ 
$0$ & $0$ & $\delta $%
\end{tabular}
\right) \psi _x-V_2\left( \psi +\int\limits_x^\infty k_{+}\psi dz\right)  
\nonumber \\
&&+\int\limits_x^\infty k_{+}\left( 
\begin{tabular}{lll}
$\alpha $ & $0$ & $0$ \\ 
$0$ & $\delta $ & $0$ \\ 
$0$ & $0$ & $\delta $%
\end{tabular}
\right) \psi _zdz  \nonumber \\
&=&\left( 
\begin{tabular}{lll}
$\alpha $ & $0$ & $0$ \\ 
$0$ & $\delta $ & $0$ \\ 
$0$ & $0$ & $\delta $%
\end{tabular}
\right) \left( \psi _x-\widehat{k}_{+}\psi \right)   \nonumber \\
&&+\left( 
\begin{tabular}{lll}
$\alpha $ & $0$ & $0$ \\ 
$0$ & $\delta $ & $0$ \\ 
$0$ & $0$ & $\delta $%
\end{tabular}
\right) \int\limits_x^\infty k_{+_x}\psi dz,  \tag{2.27}
\end{eqnarray}
where we have written $\Delta ^{\left( 2\right) }=\Delta _0^{(2)}+V_2$. On
integrating by parts in the first term, eq.$\left( 2.27\right) $ becomes 
\begin{eqnarray}
0 &=&V_2\int\limits_x^\infty k_{+}\psi dz+\int\limits_x^\infty \left( 
\begin{tabular}{lll}
$\alpha $ & $0$ & $0$ \\ 
$0$ & $\delta $ & $0$ \\ 
$0$ & $0$ & $\delta $%
\end{tabular}
\right) \widehat{k}_{+_x}\psi dz  \nonumber \\
&&+\left( V_2-\left( 
\begin{tabular}{lll}
$\alpha $ & $0$ & $0$ \\ 
$0$ & $\delta $ & $0$ \\ 
$0$ & $0$ & $\delta $%
\end{tabular}
\right) \widehat{k}_{+}\right) \psi   \nonumber \\
&&+\widehat{k}_{+}\left( 
\begin{tabular}{lll}
$\alpha $ & $0$ & $0$ \\ 
$0$ & $\delta $ & $0$ \\ 
$0$ & $0$ & $\delta $%
\end{tabular}
\right) \psi   \nonumber \\
&&+\int\limits_x^\infty \widehat{k}_{+_z}\left( 
\begin{tabular}{lll}
$\alpha $ & $0$ & $0$ \\ 
$0$ & $\delta $ & $0$ \\ 
$0$ & $0$ & $\delta $%
\end{tabular}
\right) \psi dz,  \tag{2.28}
\end{eqnarray}
and so we choose 
\begin{equation}
V_2=\left( 
\begin{tabular}{lll}
$\alpha $ & $0$ & $0$ \\ 
$0$ & $\delta $ & $0$ \\ 
$0$ & $0$ & $\delta $%
\end{tabular}
\right) \widehat{k}_{+}-\widehat{k}_{+}\left( 
\begin{tabular}{lll}
$\alpha $ & $0$ & $0$ \\ 
$0$ & $\delta $ & $0$ \\ 
$0$ & $0$ & $\delta $%
\end{tabular}
\right) ,  \tag{2.29}
\end{equation}
where $k_{+}\left( x,z;t\right) $ must satisfy 
\begin{equation}
\left( 
\begin{tabular}{lll}
$\alpha $ & $0$ & $0$ \\ 
$0$ & $\delta $ & $0$ \\ 
$0$ & $0$ & $\delta $%
\end{tabular}
\right) \widehat{k}_{+_x}+\widehat{k}_{+_z}\left( 
\begin{tabular}{lll}
$\alpha $ & $0$ & $0$ \\ 
$0$ & $\delta $ & $0$ \\ 
$0$ & $0$ & $\delta $%
\end{tabular}
\right) +V_2k_{+}=0.  \tag{2.30}
\end{equation}

We now return to the main development and describe a further important step
in the \textbf{ZS} dressing method. This is to introduce two pairs of
operators $\Delta ^{\left( i\right) }$ and $\Delta _0^{(i)}$. A typical
choice is in the following forms 
\begin{equation}
\Delta ^{\left( 1\right) }=Ii\alpha \frac \partial {\partial
t}-M;\,\,\,\,\,\,\,\,\,\Delta ^{\left( 2\right) }=I\beta \frac \partial
{\partial y}+L,\,\,\,  \tag{2.31a}
\end{equation}
and 
\begin{equation}
\Delta _0^{(1)}=Ii\alpha \frac \partial {\partial t}-M_0;\,\,\,\,\,\Delta
_0^{(2)}=I\beta \frac \partial {\partial y}+L_0,  \tag{2.31b}
\end{equation}
where $\beta $ is constant, and $M$, $L$, $M_0$, and $L_0$ are differential
operators in $x$ only. Consistent with our notation, $M_0$ and $L_0$ are
comprised of constant coefficients only and so $\Delta _0^{(1)}$, $\Delta
_0^{(2)}$ commute. Furthermore, both $\Delta _0^{(1)}$and $\Delta _0^{(2)}$
are to commute with the same operator $\Phi _F$ so that eq.$\left(
2.11\right) $ is valid. On the other hand, operators $\Delta ^{\left(
i\right) }$ are defined according to eq.$\left( 2.12\right) $, with the same 
$\Phi _{+}$. Based on this point, it is instructive to examine the operator 
\begin{equation}
P=\Delta ^{\left( 1\right) }\Delta ^{\left( 2\right) }\left( I\,+\,\Phi
_{+}\right) -\Delta ^{\left( 2\right) }\Delta ^{\left( 1\right) }\left(
I\,+\,\Phi _{+}\right) ,  \tag{2.32}
\end{equation}
which, upon the use of eq.$\left( 2.12\right) $ twice, gives 
\begin{eqnarray}
P &=&\Delta ^{\left( 1\right) }\left( I\,+\,\Phi _{+}\right) \Delta
_0^{\left( 2\right) }-\Delta ^{\left( 2\right) }\left( I\,+\,\Phi
_{+}\right) \Delta _0^{\left( 1\right) }  \nonumber \\
&=&\left( I\,+\,\Phi _{+}\right) \Delta _0^{\left( 1\right) }\Delta
_0^{\left( 2\right) }-\left( I\,+\,\Phi _{+}\right) \Delta _0^{\left(
2\right) }\Delta _0^{\left( 1\right) }  \nonumber \\
&=&\left( I\,+\,\Phi _{+}\right) \left[ \Delta _0^{\left( 1\right) },\Delta
_0^{\left( 2\right) }\right] .  \tag{2.33}
\end{eqnarray}
However, $\Delta _0^{(1)}$and $\Delta _0^{(2)}$ are chosen so that they
commute with one another; hence $P=0$. Thus we obtain 
\begin{equation}
P=\left[ \Delta ^{\left( 1\right) },\Delta ^{\left( 2\right) }\right] \left(
I\,+\,\Phi _{+}\right) =0,  \tag{2.34}
\end{equation}
and, since $\left( I\,+\,\Phi _{+}\right) $ is invertible (an easier
assumtion which can be checked in specific cases), we get 
\begin{equation}
\left[ \Delta ^{\left( 1\right) },\Delta ^{\left( 2\right) }\right] =0, 
\tag{2.35}
\end{equation}
which means that $\Delta ^{\left( 1\right) }$ commutes with $\Delta ^{\left(
2\right) }$. If we now introduce the choice given in eq.$\left( 2.31a\right) 
$ and eq.$\left( 2.31b\right) $, eq.$\left( 2.35\right) $ becomes 
\begin{eqnarray}
0 &=&\left( Ii\alpha \frac \partial {\partial t}-M\right) \left( I\beta
\frac \partial {\partial y}+L\right)   \nonumber \\
&&-\left( I\beta \frac \partial {\partial y}+L\right) \left( Ii\alpha \frac
\partial {\partial t}-M\right) ,  \tag{2.36}
\end{eqnarray}
which simplifies to 
\begin{equation}
i\alpha L_t+\beta M_y+\left[ L,M\right] =0.  \tag{2.37}
\end{equation}
This is a generalisation of Lax pair to two auxiliarly variables; the Lax
pair is recovered if we put $\beta =0$. Equation $\left( 2.37\right) $ is
the general system of the coupled \textbf{NLS} evolution equations which can
be solved by the \textbf{ZS} dressing method (\textbf{ZS} scheme). We can
also rewrite the connections among the operators $L$, $M$, $M_0$, $L_0$, $%
\Delta ^{\left( 1\right) }$ and $\Delta ^{\left( 2\right) }$ related to the
general coupled \textbf{NLS} equations as follows 
\begin{equation}
L=\Delta ^{\left( 2\right) }=L_0+V_2,  \tag{2.38a}
\end{equation}
where

\begin{equation}
L_0=\Delta _0^{\left( 2\right) },  \tag{2.38b}
\end{equation}
and

\begin{equation}
M=M_0-V_1,  \tag{2.38c}
\end{equation}
where

\begin{equation}
M_0=I\frac{\partial ^2}{\partial x^2}.  \tag{2.38d}
\end{equation}

The procedure for solving eq.$\left( 2.37\right) $ can be described in the
following explanations. The variable coefficients which arise in the '%
\textbf{dressed}' operators, $L$ and $M$, constitute the fuctions which
satisfy the system of the evolution equations. These functions are known in
terms of $k_{+}$, where $k_{+}$ is a solution of the linear integral
Marchenko equations (see eq.$\left( 2.9\right) $ and eq.$\left( 2.10\right)
) $. This equations require $F$, and $F$ is supplied by the solution of a
pair of equations (eq.$\left( 2.18\right) $), a pair since both equations $%
\left( 2.11\right) $ are to be satisfied. Note that the interesting point
here is that the eigenvalue does not appear explisitly at any stage in the 
\textbf{ZS} scheme.

\section{\textbf{The New Integrable Coupled NLS Equations}}

In this section, we investigate and derive two types of the integrable
coupled \textbf{NLS} equations with their certain perturbative terms and
also show the Lax pairs of those equations using the \textbf{ZS} dressing
method. For this purpose, we will initially choose certain matrix function $%
k_{+}$, and then apply the \textbf{ZS} dressing method provided in section 2
to derive the equations we want.

\subsection{\textbf{Type I}}

We start by choosing the following matrix function $k_{+}$ related to 
\textbf{ZS} dressing method as follows 
\begin{equation}
k_{+}=\left( 
\begin{tabular}{lll}
$\,\,h_1(x,z;t)$ & $\,\,\,\,\,\,\,\,\,\frac 1u$ & $\,\,\,\,\,\,\,\,\,\,\,%
\frac 1w$ \\ 
$\,\,\,\pm \frac 1{u^{*}}$ & $h_2\left( x,z;t\right) $ & $h_3\left(
x,z;t\right) $ \\ 
$\,\,\,\pm \frac 1{w^{*}}$ & $h_4\left( x,z;t\right) $ & $\,h_5\left(
x,z;t\right) $%
\end{tabular}
\right) ,  \tag{3.1}
\end{equation}
$\,\,\,\,\,\,\,\,\,\,\,\,\,\,\,\,\,\,\,\,\,\,\,\,\,\,\,\,\,\,\,\,\,\,\,\,\,%
\,\,\,\,\,\,\,\,\,\,\,\,\,\,\,\,\,\,\,\,\,\,\,\,\,\,\,\,\,\,\,\,\,\,\,\,\,\,%
\,\,\,\,\,\,\,\,\,\,\,\,\,\,\,\,\,\,\,\,\,\,\,\,\,\,\,\,\,\,\,\,\,\,\,\,\,\,%
\,\,\,\,\,\,\,\,\,\,\,\,\,\,\,\,\,\,\,\,\,\,\,\,\,\,\,\,\,\,\,\,\,\,\,\,\,\,%
\,\,\,\,\,\,\,\,\,\,\,\,\,\,\,\,\,\,\,\,\,\,\,\,\,\,\,\,\,\,\,\,\,\,\,\,\,\,$%
where $h_1$, $h_2$, $h_3$, $h_4$, and $h_5$ are functions which will be
calculated using eq.$\left( 2.12\right) $.

By substituting eq.$\left( 3.1\right) $ into eq.$\left( 2.24\right) $, we
obtain 
\begin{equation}
V_1\left( x,t\right) =-2\left( 
\begin{tabular}{lll}
$\,\,\,\,\,\,\,\,\,h_{1_x}$ & -$\frac{u_x}{u^2}$ & -$\frac{w_x}{w^2}$ \\ 
$\mp \frac{u_x^{*}}{\left( u^{*}\right) ^2}$ & $h_{2_x}$ & $h_{3_x}$ \\ 
$\mp \frac{w_x^{*}}{\left( w^{*}\right) ^2}$ & $h_{4_x}$ & $\,h_{5_x}$%
\end{tabular}
\right) .  \tag{3.2}
\end{equation}
On the other hand, by substituting eq.$\left( 3.1\right) $ into eq.$\left(
2.29\right) $, we find 
\begin{equation}
V_2\left( x,t\right) \,=\left( 
\begin{tabular}{lll}
$\,\,\,\,\,\,\,\,0$ & $\frac{\left( \alpha -\delta \right) }u$ & $\frac{%
\left( \alpha -\delta \right) }w$ \\ 
$\pm \frac{\left( \delta -\alpha \right) }{u^{*}}$ & $\,\,\,\,\,\,\,0$ & $%
\,\,\,\,\,\,0$ \\ 
$\pm \frac{\left( \delta -\alpha \right) }{w^{*}}$ & $\,\,\,\,\,\,\,0$ & $%
\,\,\,\,\,\,0$%
\end{tabular}
\right) .  \tag{3.3}
\end{equation}
The functions $h_{1_x}$, $h_{2_x}$, $h_{3_x}$, $h_{4_x}$ and $h_{5_x}$ in eq.%
$\left( 3.2\right) $ can be found by substituting eq.$\left( 3.3\right) $
and eq.$\left( 3.1\right) $ into eq.$\left( 2.30\right) $, 
\begin{equation}
h_{1_x}=\mp \frac{\alpha -\delta }\alpha \left( \frac{\left| u\right|
^2+\left| w\right| ^2}{\left| u\right| ^2\left| w\right| ^2}\right) , 
\tag{3.4a}
\end{equation}
\begin{equation}
h_{2_x}=\mp \frac{\delta -\alpha }\delta \left( \frac 1{\left| u\right|
^2}\right) ,  \tag{3.4b}
\end{equation}
\begin{equation}
h_{3_x}=\mp \frac{\delta -\alpha }\delta \left( \frac 1{u^{*}w}\right) , 
\tag{3.4c}
\end{equation}
\begin{equation}
h_{4_x}=\mp \frac{\delta -\alpha }\delta \left( \frac 1{uw^{*}}\right) , 
\tag{3.4d}
\end{equation}
and 
\begin{equation}
h_{5_x}=\mp \frac{\delta -\alpha }\delta \left( \frac 1{\left| w\right|
^2}\right) .  \tag{3.4e}
\end{equation}
Hence, the coupled \textbf{NLS} equation \textbf{type I} (eq.$\left(
1.1\right) $) and its complex conjugate can easily be derived by putting
equations $\left( 2.21\right) $, $\left( 2.14\right) $, $\left( 2.26\right) $%
, $\left( 3.2\right) $, and $\left( 3.3\right) $ into eq.$\left( 2.35\right) 
$.

Finally, we can obtain the Lax pair operators of eq.$\left( 1.1\right) $ and
its complex conjugate using the relations of equations $\left( 2.37\right) $
and $\left( 2.38\right) $, 
\begin{eqnarray}
L &=&\left( 
\begin{tabular}{lll}
$\alpha $ & $0$ & $0$ \\ 
$0$ & $\delta $ & $0$ \\ 
$0$ & $0$ & $\delta $%
\end{tabular}
\right) \frac \partial {\partial x}  \nonumber \\
&&+\left( 
\begin{tabular}{lll}
$\,\,\,\,\,\,\,\,0$ & $\frac{\left( \alpha -\delta \right) }u$ & $\frac{%
\left( \alpha -\delta \right) }w$ \\ 
$\pm \frac{\left( \delta -\alpha \right) }{u^{*}}$ & $\,\,\,\,\,\,\,0$ & $%
\,\,\,\,\,\,0$ \\ 
$\pm \frac{\left( \delta -\alpha \right) }{w^{*}}$ & $\,\,\,\,\,\,\,0$ & $%
\,\,\,\,\,\,0$%
\end{tabular}
\right) ,  \tag{3.5a}
\end{eqnarray}
and 
\begin{equation}
M=I\frac{\partial ^2}{\partial x^2}+2\left( 
\begin{tabular}{lll}
$\,\,\,\,\,\,\,\,\,h_{1_x}$ & -$\frac{u_x}{u^2}$ & -$\frac{w_x}{w^2}$ \\ 
$\mp \frac{u_x^{*}}{\left( u^{*}\right) ^2}$ & $h_{2_x}$ & $h_{3_x}$ \\ 
$\mp \frac{w_x^{*}}{\left( w^{*}\right) ^2}$ & $h_{4_x}$ & $\,h_{5_x}$%
\end{tabular}
\right) .  \tag{3.5b}
\end{equation}

\subsection{\textbf{Type II}}

Following the same procedure as type I, we only choose $k_{+}$ in a
different form from that in eq.$\left( 3.1\right) $, 
\begin{equation}
k_{+}=\,\left( 
\begin{tabular}{lll}
$\,a(x,z;t)$ & $\,\,\,\,\,\,\,\,\frac u\aleph $ & $\,\,\,\,\,\,\,\,\frac
w\aleph $ \\ 
$\,\frac{\pm u^{*}}\aleph $ & $\,d\left( x,z;t\right) $ & $\,e\left(
x,z;t\right) $ \\ 
$\,\,\,\frac{\pm w^{*}}\aleph $ & $f\left( x,z;t\right) $ & $\,g\left(
x,z;t\right) $%
\end{tabular}
\right) ,  \tag{3.6}
\end{equation}
$\,\,\,\,\,\,\,\,\,\,\,\,\,\,\,\,\,\,\,\,\,\,\,\,\,\,\,\,\,\,\,\,\,\,\,\,\,%
\,\,\,\,\,\,\,\,\,\,\,\,\,\,\,\,\,\,\,\,\,\,\,\,\,\,\,\,\,\,\,\,\,\,\,\,\,\,%
\,\,\,\,\,\,\,\,\,\,\,\,\,\,\,\,\,\,\,\,\,\,\,\,\,\,\,\,\,\,\,\,\,\,\,\,\,\,%
\,\,\,\,\,\,\,\,\,\,\,\,\,\,\,\,\,\,\,\,\,\,\,\,\,\,\,\,\,\,\,\,\,\,\,\,\,\,%
\,\,\,\,\,\,\,\,\,\,\,\,\,\,\,\,\,\,\,\,\,\,\,\,\,\,\,\,\,\,\,\,\,\,\,\,\,\,%
\,\,\,\,\,$where $a$, $d$, $e$, $f$ and $g$ are are functions which will be
investigated using eq.$\left( 2.12\right) $ and $\aleph \left( x,z;t\right) =%
\sqrt{1+s\left( \left| u\right| ^2+\left| w\right| ^2\right) }$.

By substituting eq.$\left( 3.6\right) $ into eq.$\left( 2.24\right) $, we
obtain 
\begin{equation}
V_1\left( x,t\right) =-2\left( 
\begin{tabular}{lll}
$\,\,\,\,\,\,\,a_x$ & $\left( \frac u\aleph \right) _x$ & $\left( \frac
w\aleph \right) _x$ \\ 
$\left( \frac{\pm u^{*}}\aleph \right) _x$ & $\,\,\,\,\,\,\,d_x$ & $%
\,\,\,\,\,\,\,\,e_x$ \\ 
$\left( \frac{\pm w^{*}}\aleph \right) _x$ & $\,\,\,\,\,\,\,\,f_x$ & $%
\,\,\,\,\,\,\,g_x$%
\end{tabular}
\right) .  \tag{3.7}
\end{equation}
On the other hand, by substituting eq.$\left( 3.6\right) $ into eq.$\left(
2.29\right) $, we find 
\begin{equation}
V_2\left( x,t\right) \,=\left( 
\begin{tabular}{lll}
$\,\,\,\,\,\,\,\,\,0$ & $\frac{\left( \alpha -\delta \right) u}\aleph $ & $%
\frac{\left( \alpha -\delta \right) w}\aleph $ \\ 
$\frac{\pm \left( \delta -\alpha \right) u^{*}}\aleph $ & $\,\,\,\,\,\,\,0$
& $\,\,\,\,\,\,\,0$ \\ 
$\frac{\pm \left( \delta -\alpha \right) w^{*}}\aleph $ & $\,\,\,\,\,\,\,0$
& $\,\,\,\,\,\,\,0$%
\end{tabular}
\right) .  \tag{3.8}
\end{equation}
The functions $a_x$, $d_x$, $e_x$, $f_x$ and $g_x$ in eq.$\left( 3.7\right) $
can be found by substituting eq.$\left( 3.8\right) $ and eq.$\left(
3.6\right) $ into eq.$\left( 2.30\right) $, 
\begin{equation}
a_{_x}=\mp \frac{\alpha -\delta }\alpha \left( \frac{\left| u\right|
^2+\left| w\right| ^2}{1+s\left( \left| u\right| ^2+\left| w\right|
^2\right) }\right) ,  \tag{3.9a}
\end{equation}
\begin{equation}
d_{_x}=\mp \frac{\delta -\alpha }\delta \left( \frac{\left| u\right| ^2}{%
1+s\left( \left| u\right| ^2+\left| w\right| ^2\right) }\right) ,  \tag{3.9b}
\end{equation}
\begin{equation}
e_{_x}=\mp \frac{\delta -\alpha }\delta \left( \frac{u^{*}w}{1+s\left(
\left| u\right| ^2+\left| w\right| ^2\right) }\right) ,  \tag{3.9c}
\end{equation}
\begin{equation}
f_{_x}=\mp \frac{\delta -\alpha }\delta \left( \frac{uw^{*}}{1+s\left(
\left| u\right| ^2+\left| w\right| ^2\right) }\right) ,  \tag{3.9d}
\end{equation}
and 
\begin{equation}
g_{_x}=\mp \frac{\delta -\alpha }\delta \left( \frac{\left| w\right| ^2}{%
1+s\left( \left| u\right| ^2+\left| w\right| ^2\right) }\right) .  \tag{3.9e}
\end{equation}
Hence, the coupled \textbf{NLS} equation \textbf{type II} (eq.$\left(
1.3\right) $) and its complex conjugate can be derived by substituting
equations $\left( 2.21\right) $, $\left( 2.14\right) $, $\left( 2.26\right) $%
, $\left( 3.7\right) $, and $\left( 3.8\right) $ into eq.$\left( 2.35\right) 
$.

We can then obtain the Lax pair operators of eq.$\left( 1.1\right) $ and its
complex conjugate using the relations of equations $\left( 2.37\right) $ and$%
\,\left( 2.38\right) $, 
\begin{eqnarray*}
L &=&\left( 
\begin{tabular}{lll}
$\alpha $ & $0$ & $0$ \\ 
$0$ & $\delta $ & $0$ \\ 
$0$ & $0$ & $\delta $%
\end{tabular}
\right) \frac \partial {\partial x} \\
&&+\left( 
\begin{tabular}{lll}
$\,\,\,\,\,\,\,\,\,0$ & $\frac{\left( \alpha -\delta \right) u}\aleph $ & $%
\frac{\left( \alpha -\delta \right) w}\aleph $ \\ 
$\frac{\pm \left( \delta -\alpha \right) u^{*}}\aleph $ & $\,\,\,\,\,\,\,0$
& $\,\,\,\,\,\,\,0$ \\ 
$\frac{\pm \left( \delta -\alpha \right) w^{*}}\aleph $ & $\,\,\,\,\,\,\,0$
& $\,\,\,\,\,\,\,0$%
\end{tabular}
\right) ,
\end{eqnarray*}
\begin{equation}
\tag{3.10a}
\end{equation}
and 
\begin{equation}
M=I\frac{\partial ^2}{\partial x^2}+2\left( 
\begin{tabular}{lll}
$\,\,\,\,\,\,\,a_x$ & $\left( \frac u\aleph \right) _x$ & $\left( \frac
w\aleph \right) _x$ \\ 
$\left( \frac{\pm u^{*}}\aleph \right) _x$ & $\,\,\,\,\,\,\,d_x$ & $%
\,\,\,\,\,\,\,\,e_x$ \\ 
$\left( \frac{\pm w^{*}}\aleph \right) _x$ & $\,\,\,\,\,\,\,\,f_x$ & $%
\,\,\,\,\,\,\,g_x$%
\end{tabular}
\right) .  \tag{3.10b}
\end{equation}

\section{\textbf{The solutions of the integrable coupled NLS equations}}

In this section, we investigate the solutions of both coupled \textbf{NLS}
equations (type I and type II). We firstly choose the general matrix
function $F$, 
\begin{equation}
F\,=\,\left( 
\begin{tabular}{lll}
$\,\,\,\,\,\,\,\,\,\,\,\,0$ & $A\left( x,z;t\right) $ & $B\left(
x,z;t\right) $ \\ 
$A^{*}\left( x,z;t\right) $ & $\,\,\,\,\,\,\,\,\,\,\,\,0$ & $%
\,\,\,\,\,\,\,\,\,\,\,\,\,0$ \\ 
$B^{*}\left( x,z;t\right) $ & $\,\,\,\,\,\,\,\,\,\,\,\,0$ & $%
\,\,\,\,\,\,\,\,\,\,\,\,\,0$%
\end{tabular}
\right) ,  \tag{4.1}
\end{equation}
where the functions $A\left( x,z;t\right) $, $B\left( x,z;t\right) $ and
their complex conjugate will be found by solving the linear differential
equation in eq.$\left( 2.18\right) $. Hence, after substituting eq.$\left(
4.1\right) $ into eq.$\left( 2.18\right) $, we get the following four linear
differential equations 
\begin{eqnarray}
i\alpha A_t\,+A_{zz}\,-\,A_{xx}\, &=&\,0,  \tag{4.2a} \\
i\alpha B_t\,+B_{zz}\,-\,B_{xx}\, &=&\,0,  \tag{4.2b} \\
i\alpha A_t^{*}\,\,+A_{zz}^{*}\,-\,A_{xx}^{*}\, &=&\,0,  \tag{4.2c} \\
i\alpha B_t^{*}\,+\,B_{zz}^{*}\,-\,B_{xx}^{*}\, &=&\,0\,.  \tag{4.2d}
\end{eqnarray}
$\,$The solution of the above$\,\,$equations$\,\,$can be derived by using
separable variable method. We then find 
\begin{eqnarray}
A\, &=&\,A_0e^{-\alpha \rho z}\left[ e^{\rho \left( \delta x+i\rho \left(
\alpha ^2-\delta ^2\right) t\right) }\right] ,  \tag{4.3a} \\
B &=&B_0e^{-\alpha \rho z}\left[ e^{\rho \left( \delta x+i\rho \left( \alpha
^2-\delta ^2\right) t\right) }\right] ,  \tag{4.3b} \\
A^{*} &=&A_0^{*}e^{-\delta \sigma z}\left[ e^{\sigma \left( \alpha x+i\sigma
\left( \delta ^2-\alpha ^2\right) t\right) }\right] ,  \tag{4.3c} \\
B^{*}\, &=&\,B_0^{*}e^{-\delta \sigma z}\left[ e^{\sigma \left( \alpha
x+i\sigma \left( \delta ^2-\alpha ^2\right) t\right) }\right] ,  \tag{4.3d}
\end{eqnarray}
where $A_0\,$\thinspace , $A_0^{*}$, $B_0\,$\thinspace and\thinspace $%
B_0^{*}\,\,$are arbitrary complex parameters, and $\rho $ and $\sigma $ are
chosen as imaginary constants in which $\rho ^{*}=-\sigma $.

\subsection{\textbf{The solution of the type I}}

Here, we derive the solution of the coupled \textbf{NLS} equation type I
using the \textbf{ZS} dressing method explained in section 2. By putting eq.$%
\left( 3.1\right) $ and eq.$\left( 4.1\right) $ into eq.$\left( 2.9\right) $%
, we get (for $h_1$, $u_{\text{,}}\,$and $w$) 
\[
h_1\,=\,-\int\limits_x^\infty \,\frac{A_0^{*}}ue^{i\frac{\sigma ^2}\alpha
\left( \delta ^2-\alpha ^2\right) t}e^{-\delta \sigma z}e^{\alpha \sigma
y}dy\,\,\,\,\,\,\,\,\,\,\,\,\,\,\,\,\,\,\,\,\,\,\,\,\,\,\,\,\,\,\,\,\,\,\,\,%
\,\,\,\,\,\,\,\, 
\]
\begin{equation}
-\int\limits_x^\infty \frac{B_0^{*}}we^{i\frac{\sigma ^2}\alpha \left(
\delta ^2-\alpha ^2\right) t}e^{-\delta \sigma z}e^{\alpha \sigma y}dy, 
\tag{4.4a}
\end{equation}
\begin{eqnarray*}
\frac 1u\, &=&\,-\left( e^{-\alpha \rho z}A_0\right) e^{\delta \rho x}e^{i%
\frac{\rho ^2}\alpha \left( \alpha ^2-\delta ^2\right) t} \\
&&\,-\left( A_0\right) \int\limits_x^\infty h_1\,e^{i\frac{\rho ^2}\alpha
\left( \alpha ^2-\delta ^2\right) t\,}e^{-\alpha \rho z}e^{\delta \rho
y}dy\,,
\end{eqnarray*}
\begin{equation}
\tag{4.4b}
\end{equation}
and 
\begin{eqnarray*}
\frac 1w &=&-\left( e^{-\alpha \rho z}B_0\right) e^{\delta \rho x}e^{i\frac{%
\rho ^2}\alpha \left( \alpha ^2-\delta ^2\right) t} \\
&&-\left( B_0\right) \int\limits_x^\infty h_1e^{i\frac{\rho ^2}\alpha \left(
\alpha ^2-\delta ^2\right) t}e^{-\alpha \rho z}e^{\delta \rho y}dy.
\end{eqnarray*}
\begin{equation}
\tag{4.4c}
\end{equation}
The final solution is work on $x=z$ as the \textbf{ZS} method works (see the
explanation of eq.$\left( 2.11\right) $ in section 2.2). Hence, by
substituting eq.$(4.4a)$ to eq.$(4.4b)\,$and eq.$(4.4c)$ we find the
solution of the type I: 
\begin{equation}
u\left( x,t\right) \,=-\frac{1\,+\,\Im \Box _1\Box _2}{A_0\,\Box _1}\,\,, 
\tag{4.5a}
\end{equation}
and 
\begin{equation}
w\left( x,t\right) =-\frac{1\,+\,\Im \Box _1\Box _2}{B_0\,\Box _1}, 
\tag{4.5b}
\end{equation}
where 
\begin{equation}
\Im =\frac{\mu \left( \left| A_0\right| ^2+\left| B_0\right| ^2\right) }{%
\left( \left( \frac{\alpha ^2-\delta ^2}\alpha \right) ^{1/2}\left( \rho
-\sigma \right) \right) ^2},  \tag{4.6a}
\end{equation}
\begin{equation}
\Theta _{11}=e^{\rho \left( \delta -\alpha \right) x},  \tag{4.6b}
\end{equation}
\begin{equation}
\Theta _{21}=e^{-\sigma \left( \delta -\alpha \right) x},  \tag{4.6c}
\end{equation}
\begin{equation}
\Theta _{12}=e^{-i\frac{\rho ^2}\alpha \left( \delta ^2-\alpha ^2\right) t},
\tag{4.6d}
\end{equation}
\begin{equation}
\Theta _{22}=e^{i\frac{\sigma ^2}\alpha \left( \delta ^2-\alpha ^2\right) t},
\tag{4.6e}
\end{equation}
\begin{equation}
\Box _1=\Theta _{11}\Theta _{12},  \tag{4.6f}
\end{equation}
and 
\begin{equation}
\Box _2=\Theta _{21}\Theta _{22}.  \tag{4.6g}
\end{equation}
So, we can clearly see that the solution of the type I has no soliton
solution.

\subsection{\textbf{The Solution of the Type II\thinspace \thinspace }$\,$}

By reviewing the solution's steps in section 4.1, we can also investigate
the solution of the type II. We then start by substituting eq.$\left(
3.6\right) $ and eq.$\left( 4.1\right) $ into eq.$\left( 2.9\right) $, we
get (for $a$, $u_{\text{,}}\,$and $w$) 
\[
a\,=\,-\int\limits_x^\infty \,\frac{uA_0^{*}}\aleph e^{i\frac{\sigma ^2}%
\alpha \left( \delta ^2-\alpha ^2\right) t}e^{-\delta \sigma z}e^{\alpha
\sigma
y}dy\,\,\,\,\,\,\,\,\,\,\,\,\,\,\,\,\,\,\,\,\,\,\,\,\,\,\,\,\,\,\,\,\,\,\,\,%
\,\,\,\,\,\,\,\,
\]
\begin{equation}
-\int\limits_x^\infty \frac{wB_0^{*}}\aleph e^{i\frac{\sigma ^2}\alpha
\left( \delta ^2-\alpha ^2\right) t}e^{-\delta \sigma z}e^{\alpha \sigma
y}dy,  \tag{4.7a}
\end{equation}
\begin{eqnarray*}
\frac u\aleph  &=&-\left( e^{-\alpha \rho z}A_0\right) e^{\delta \rho x}e^{i%
\frac{\rho ^2}\alpha \left( \alpha ^2-\delta ^2\right) t} \\
&&-A_0\int\limits_x^\infty a\,e^{i\frac{\rho ^2}\alpha \left( \alpha
^2-\delta ^2\right) t\,}e^{-\alpha \rho z}e^{\delta \rho
y}dy\,,\,\,\,\,\,\,\,\,\,\,\,\,\,\,\,\,\,\,\,\,\,\,\,\,\,\,\,\,\,\,\,\,\,\,%
\,\,\,\,\,\,\,\,\,\,
\end{eqnarray*}
\begin{equation}
\tag{4.7b}
\end{equation}
and 
\begin{eqnarray*}
w &=&-\aleph \left( e^{-\alpha \rho z}B_0\right) e^{\delta \rho x}e^{i\frac{%
\rho ^2}\alpha \left( \alpha ^2-\delta ^2\right) t} \\
&&-\aleph \left( B_0\right) \int\limits_x^\infty ae^{i\frac{\rho ^2}\alpha
\left( \alpha ^2-\delta ^2\right) t}e^{-\alpha \rho z}e^{\delta \rho y}dy.
\end{eqnarray*}
\begin{equation}
\tag{4.7c}
\end{equation}
Hence, by substituting eq.$(4.7a)$ to eq.$(4.7b)\,$and eq.$(4.7c)$ we find
the solution of the type II: 
\begin{equation}
u\left( x,t\right) \,=-\left( \frac{1+s\Diamond _1}{1-s\Upsilon _1}\right)
^{\frac 12}\left( \frac{A_0\,\Box _1}{1\,+\,\Im \Box _1\Box _2}\right) \,\,,
\tag{4.8a}
\end{equation}
and 
\begin{equation}
w\left( x,t\right) =-\left( \frac{1+s\Diamond _2}{1-s\Upsilon _2}\right)
^{\frac 12}\left( \frac{B_0\,\Box _1}{1\,+\,\Im \Box _1\Box _2}\right) , 
\tag{4.8b}
\end{equation}
where 
\begin{equation}
\Upsilon _1=\frac{\left| A_0\right| ^2\Box _1\Box _1^{\prime }}\Xi , 
\tag{4.9a}
\end{equation}
\begin{equation}
\Upsilon _2=\frac{\left| B_0\right| ^2\Box _1\Box _1^{\prime }}\Xi , 
\tag{4.9b}
\end{equation}
\[
\Xi =1+\Im \Box _1\Box _2+\Im \Box _1^{\prime }\Box _2^{\prime }+\Im ^2\Box
_1\Box _2\Box _1^{\prime }\Box _2^{\prime },
\]
\begin{equation}
\tag{4.9c}
\end{equation}
\begin{equation}
\Diamond _1=\frac{\Upsilon _1}{\left( 1-\frac{s\Upsilon _1}{1-s\Upsilon _2}%
\right) \left( 1-s\Upsilon _2\right) },  \tag{4.9d}
\end{equation}
\begin{equation}
\Diamond _2=\frac{\Upsilon _2}{\left( 1-\frac{s\Upsilon _2}{1-s\Upsilon _1}%
\right) \left( 1-s\Upsilon _1\right) },  \tag{4.9e}
\end{equation}
\begin{equation}
\Theta _{11}^{\prime }=e^{\rho ^{*}\left( \delta -\alpha \right) x}, 
\tag{4.9f}
\end{equation}
\begin{equation}
\Theta _{21}^{\prime }=e^{-\sigma ^{*}\left( \delta -\alpha \right) x}, 
\tag{4.9g}
\end{equation}
\begin{equation}
\Theta _{12}^{\prime }=e^{i\frac{\left( \rho ^2\right) ^{*}}\alpha \left(
\delta ^2-\alpha ^2\right) t},  \tag{4.9h}
\end{equation}
\begin{equation}
\Theta _{22}^{\prime }=e^{-i\frac{\left( \sigma ^2\right) ^{*}}\alpha \left(
\delta ^2-\alpha ^2\right) t},  \tag{4.9i}
\end{equation}
\begin{equation}
\Box _1^{\prime }=\Theta _{11}^{\prime }\Theta _{12}^{\prime },  \tag{4.9j}
\end{equation}
and 
\begin{equation}
\Box _2^{\prime }=\Theta _{21}^{\prime }\Theta _{22}^{\prime }.  \tag{4.9k}
\end{equation}

It seems that the solution of the type II in equations $\left( 4.8a\right) $
and $\left( 4.8b\right) $ can roughly be identified as \textbf{a stable
soliton solution} when parameter $s$ is small. If we put $s\rightarrow 0$
either in eq.$\left( 1.6\right) $ or in equations $\left( 4.8\right) $ then
we will find the one soliton solution of the coupled \textbf{NLS} equation
of Manakov type appeared in ref.$\left[ 6\right] $.

We can also choose parameters $\alpha =4.5$ and $\delta =1.73$ in our type
II equation and its solutions, respectively. The aim of this choice is to
make parameters $\chi =\mu =\frac 12$and to reduce our one soliton (in eq.$%
\left( 4.8a\right) $ and eq.$\left( 4.8b\right) $) to be a simple solution.

It is very interesting to expand the discussions related to an equation
which has a soliton solution. So, it is undoubtedly that the type II
equation is particularly to be discussed in this paper. Even the equation $%
\left( 1.3\right) $ with a simple additional term has experimentally been
investigated by Christodoulides et al. $\left[ 7\right] $, and Kivshar et
al. $\left[ 8\right] $. Based on their expanations, we suppose that there
must be a physical meaning of our type II equation closely related to that
in ref.$\left[ 7,8\right] .$

\section{\textbf{Concluding Remarks}}

In this paper, we have presented two types of the integrable coupled \textbf{%
NLS} equation with perturbative terms derived using the \textbf{ZS} dressing
method and have also found their solutions. Using the Lax pairs of the two
types, we have identified the existence of the integrability of those
equations. The very important step to obtain the two types of the coupled 
\textbf{NLS} equations in the \textbf{ZS} dressing method is the choice of $%
k_{+}$ in eq.$\left( 3.1\right) $ and eq.$\left( 3.6\right) $, respectively.
The type I has no soliton solution. However, the type II could exactly have
a stable soliton solution when the effective saturation parameter, $s$ is
small $\left( s\rightarrow 0\right) $.

In the two types, the coupled \textbf{NLS} equations maybe have exclusively
physical meanings. They may describe wave propagations in birefringent
optical fibers with nonlinear effects such as the Kerr effect, The Raman
Scattering and gain and loss effects.

The type II is very interesting to be discussed because it has a complicated
soliton solution . We also propose that this type must also have
multisolitons solutions. However, the cases will be disscussed in our next
investigations $\left[ 10\right] $.

$\mathbf{ACKNOWLEDGEMENTS}$

Author would like to thank Dr. N.N. Akhmediev\textbf{\ }(Optical Sciences
Centre, The Institute for Advanced Studies, The Australian National
University ) for useful discussions. We also would like to thank Dr. F.P.
Zen \thinspace for his encouragements.The work of H.E. is supported by 
\textbf{CIDA - EIUDP}, Republic of Indonesia.

\end{document}